\documentclass[superscriptaddress,showpacs,aps,footinbib,twocolumn,prl,floatfix]{revtex4-1}

\usepackage{graphicx,subfigure,amsmath,bbm}
\usepackage{graphicx}
\usepackage{dcolumn}
\usepackage{bm}
\usepackage{amsmath}
\usepackage{amsthm}

\usepackage[english]{babel}
\usepackage[utf8]{inputenc}
\usepackage{graphicx}
\usepackage[colorinlistoftodos]{todonotes}
\usepackage{bbold }
\usepackage{physics}
\usepackage[utf8]{inputenc}
\usepackage{amssymb }
\usepackage{graphicx}
\DeclareGraphicsExtensions{.eps}
\usepackage{epstopdf}
\usepackage{latexsym}
\usepackage{amsfonts}
\usepackage{xcolor}
\usepackage{units}
\usepackage{bm}
\usepackage{verbatim}
\usepackage[colorlinks = true,
            linkcolor = red,
            urlcolor  = blue,
            citecolor = magenta,
            anchorcolor = red]{hyperref}
\usepackage{esint}
\usepackage{soul}
\usepackage{cancel}
\usepackage[normalem]{ulem}
\usepackage{ dsfont }
\usepackage{empheq}
\usepackage{ wasysym }
\usepackage{scalerel}

\DeclareMathOperator{\arccosh}{arcCosh}

\definecolor{AB-color}{RGB}{128,0,128}

\definecolor{GB-color}{RGB}{20,100,30}

\newcommand{\uu}{\mathrm{u}}
\newcommand{\vv}{\mathrm{v}}

\fontsize{11.4}{16} \selectfont

\begin{document}

\title{Nonlocal Thermoelectricity in a\\ Superconductor–Topological-Insulator–Superconductor Junction\\ in Contact with a Normal-Metal Probe: Evidence for Helical Edge States}

\author{Gianmichele Blasi}
\email{gianmichele.blasi@sns.it} 
\affiliation{NEST, Scuola Normale Superiore and Instituto Nanoscienze-CNR, I-56126, Pisa, Italy}
\author{Fabio Taddei}
\affiliation{NEST, Scuola Normale Superiore and Instituto Nanoscienze-CNR, I-56126, Pisa, Italy}
\author{Liliana Arrachea}
\affiliation{International Center for Advanced Studies, ECyT-UNSAM, Campus Miguelete, 25 de Mayo y Francia, 1650 Buenos Aires, Argentina}
\author{Matteo Carrega}
\affiliation{NEST, Scuola Normale Superiore and Instituto Nanoscienze-CNR, I-56126, Pisa, Italy}
\author{Alessandro Braggio}
\email{alessandro.braggio@nano.cnr.it} 
\affiliation{NEST, Scuola Normale Superiore and Instituto Nanoscienze-CNR, I-56126, Pisa, Italy}
\begin{abstract}
We consider a Josephson junction hosting a Kramers pair of helical edge states of a quantum spin Hall bar in contact with a normal-metal probe. In this hybrid system, the orbital phase induced by a small magnetic field threading the junction known as Doppler shift (DS), combines with the conventional Josephson phase difference and originates an effect akin to a Zeeman field in the spectrum.
As a consequence, when a temperature bias is applied to the superconducting terminals, a thermoelectric current is established in the normal probe. We argue that this purely non-local thermoelectric effect is a unique signature of  the helical nature of the edge states coupled to superconducting leads and it can constitute a useful tool for probing the helical nature of the edge states in systems where the Hall bar configuration is difficult to achieve.
We fully characterize thermoelectric response and performance of this hybrid junction in a wide range of parameters, demonstrating that the external magnetic flux inducing the DS can be used as a knob to control the thermoelectric response and the heat flow in a novel device based on topological junctions.
\end{abstract}
\maketitle

{\em Introduction.}--- 
Quantum spin Hall systems in two-dimensional topological insulators (TI)
are receiving a lot of attention~\cite{moore2009,hasan2010,Qi2010,ando2013} due to their non-trivial topological properties. 
 The clearest signature of the quantum spin Hall phase is the existence of Kramers pairs of helical edge states, which propagate in opposite directions with opposite spin orientations (spin-momentum locking)~\cite{Tkachov_book}.
After the pioneering theoretical ideas \cite{ti1,ti2,ti3} and experimental realizations in HgTe quantum wells \cite{ti4,ti5,ti6}, other platforms to realize this topological phase, preserving time-reversal symmetry,
have been proposed in different materials \cite{cob,wu,tang,jia,claessen1,claessen2,fuhrer}.
In HgTe the helical nature of the edge states is commonly
probed by means of nonlocal transport measurements in a Hall bar geometry with four or more terminals~\cite{ti4,ti5,ti6} and quantum point contacts~\cite{citro2011,ronetti2016,ronetti2017}. This can be very hard to implement in some other systems, where evidence is shown on the existence of edge states but not yet on their helical nature \cite{cob,wu,tang,jia,claessen1,claessen2,fuhrer}.

When the Kramers pairs of helical edge states are embedded in a superconducting junction, the 
Andreev states inherit non-trivial properties.
Topological Josephson junctions formed by two-dimensional TIs have been studied recently~\cite{teoj1,barbarino2013,teoj2,teoj3,teoj4,teoj5,teoj6,kopnin,tkachovhan,ther1,ther2,ther3,ther4,ther5} and experimentally realized~\cite{expj0,expj01,expj1,expj2}.
In particular, a small magnetic flux in topological junctions can lead to very interesting features due to the effective orbital {\em Doppler shift} (DS) acquired 
by the electrons in the edge states \cite{teoj6}. 

\begin{figure}[ht!]
	\centering
	\includegraphics[width=\columnwidth]{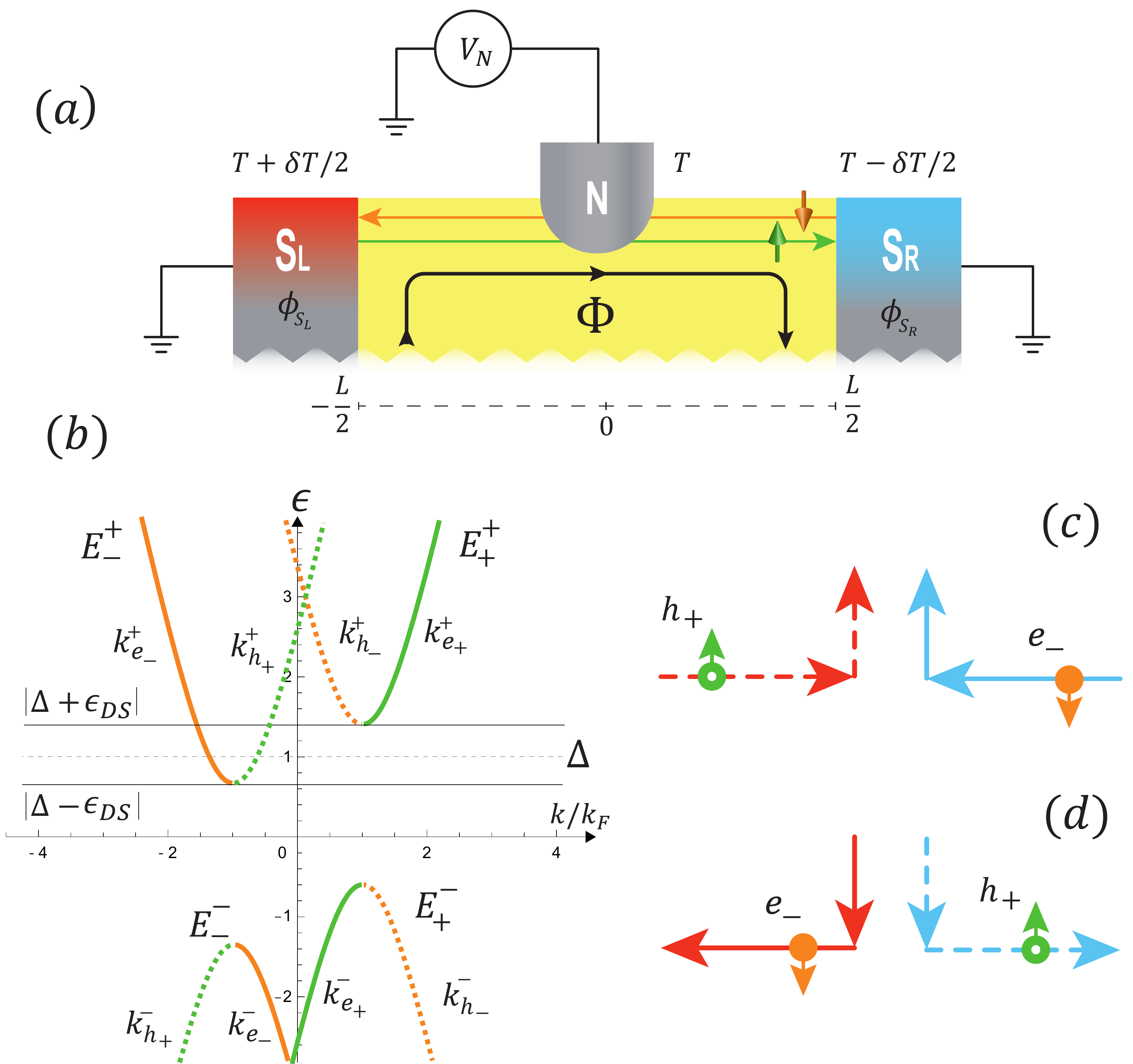}
	\caption{(a) Sketch of the setup. A helical Kramers pair of edge states of the quantum spin Hall effect is contacted by two superconductors at different temperatures, $T_{\rm{S_L}}=T+\delta T/2$ and
	$T_{\rm{S_R}}=T-\delta T/2$ and with a normal-metal probe at temperature $T_N=T$ at which a bias voltage $V_N$ is possibly applied. The structure is threaded by a magnetic flux which induces a Doppler shift in the edge states in addition to a Josephson phase difference applied between the two superconductors. (b) Dispersion curves for  quasiparticles $e_{\pm}$ (solid lines) and quasiholes $h_{\pm}$ (dashed lines) in the proximized superconductor $\rm{S_{L}/S_{R}}$ for $0 <\epsilon_{DS}(\Phi) <\Delta$.  Transport processes are depicted in panel $(c)$ for $V_N=0, \delta T\neq 0$ and in panel $(d)$ for $V_N\neq 0, \delta T= 0$,  when  the spectrum for $e_+,h_-$ is assumed fully gapped. }
	\label{fig:fig1}
\end{figure}

In the present work we argue that the DS leads to a nonlocal thermoelectric effect as a unique consequence of the helical nature of the edge states. 
The setup under investigation is shown in Fig.~\ref{fig:fig1}$(a)$, where a pair of edge states are contacted to superconductors,
while  a normal-metal probe -- such as STM tip~\cite{
das2011,liu2015,hus2017,voigtlander2018} -- is directly contacted to the edge states.  
A similar geometry has been considered in Refs.~\onlinecite{ther3,ther2}, where however, only the voltage-driven charge transport in the tunneling regime or the heat transport with no probe have been analyzed. 
In the absence of 
DS, particle-hole symmetry, inherently present in superconducting systems, prevents the development of 
any thermoelectric effect. Remarkably, the DS has an effect akin to a Zeeman splitting in the two spin-polarized
 members of the Kramers pair. 
 Although the whole system is particle-hole symmetric by construction, the local density of states for each spin species, at the contact with the probe, lacks the symmetry between positive and negative energies due to the DS. 
Therefore, when a temperature difference is applied between the two superconductors a thermoelectric current flows between the TI and the probe. 
The key for this response is the fact that the proximity to superconductors gives rise to a simultaneous flow of helical electrons and holes. Since they move in opposite directions, they thermalize with different reservoirs, see Fig.~\ref{fig:fig1}$(c)$.
The intrinsic particle-hole symmetry of a normal metal-superconducting junction can be broken, in order to generate thermoelectric current, by using a Zeeman field and spin-polarized barriers or nonlinearities~\cite{seba1,seba2,keidel,marchegiani}.
Our proposal, on the contrary, relies on a completely different mechanisms which makes use of the helicity of the edge states under the effect of the DS. In the following we quantitatively discuss this peculiar effect in the linear response regime, using the scattering matrix approach. We analyze different figures of merit, and show that it is possible to achieve very high values of the nonlocal Seebeck coefficient. 

{\em Model.}--- We consider the topological Josephson junction depicted in Fig.~\ref{fig:fig1}$(a)$ with the upper edge of length $L$ tunnel coupled with a normal (N) probe. The width of the TI is assumed large such that upper and lower edges are decoupled, and thus we focus only on the former one.  
The two electrodes induce superconducting correlations on the edge states via proximity effect~\cite{teoj6,ther1}.
The associated Bogoliubov-de Gennes (BdG) Hamiltonian reads
\begin{equation}
\label{My_Hamiltonian}
 {\cal H}=\mqty(H(x) & i\sigma_y\Delta (x) \\ -i\sigma_y\Delta (x)^*  &  -H(x)^*) ,
\end{equation}
expressed in the four-component Nambu basis $(\psi_{\uparrow},\psi_{\downarrow},\psi_{\uparrow}^{*},\psi_{\downarrow}^{*})^T$ with spin $\uparrow$ and $\downarrow$ collinear with natural spin quantization axis of the TI edge pointing along $z$-direction, and where $H(x)=v_F\left(-i\hbar\partial_x+p_{DS}/2\right)\sigma_z-\mu\sigma_0+\Lambda(x)$ with $-H(x)^*$ its time-reversal partner. We include also a contact potential $\Lambda(x)=\Lambda\delta(x+x_0)+\Lambda\delta(x-x_0) $ at the boundaries with $x_0=L/2$; $v_F$ indicates Fermi velocity, $\mu$ is the chemical potential and $\sigma_i$ are the Pauli matrices.
The momentum $p_{DS}=(\pi \hbar/L)(\Phi/\Phi_0)$ represents the so-called doppler shift (DS) contribution describing the gauge invariant shift of momentum induced by a small magnetic flux $\Phi$ through the TI junction~\cite{teoj6}, while
$\Phi_0=h/2e$ is the magnetic flux quantum. We consider rigid boundary conditions for the order parameter $\Delta(x) = \Delta\left[\Theta(-x -L/2)e^{i\phi_{S_L}} + \Theta(x - L/2)e^{i\phi_{S_R}}\right]$, with $\Theta(x)$ the step function and $\phi_{S_L}$, $\phi_{S_R}$ the superconducting phase in the left/right superconductor, with an induced gap amplitude $\Delta$ due to proximization. 

The eigenspectrum of the BdG Hamiltonian relative to the homogeneous proximized TI edge, is reported in Fig.~\ref{fig:fig1}$(b)$ and is given by $E^j_{\pm}(k)=\left(\epsilon_{DS}(\Phi)+j\sqrt{(\hbar v_F k\mp \mu)^2+\Delta^2}\right)$, with $j=\pm$ indicating branches with positive/negative concavity and $\epsilon_{DS}(\Phi)=v_F p_{DS}/2=(v_F h/4 L)(\Phi/\Phi_0)$ being the Doppler-shift energy.
The effect of the DS on the dispersion curve is to shift the various branches vertically by an amount $\epsilon_{DS}(\Phi)$, upwards or downwards, as shown in Fig.~\ref{fig:fig1}$(b)$. A finite value of the magnetic flux $\Phi$ reduces  the gap, which closes when $\abs{\epsilon_{DS}(\Phi)}=\Delta$.
The quasiparticle (QP) eigenfunctions in Nambu notation of both left/right superconductors ($i=\rm{S_L,S_R}$) are given by
\begin{eqnarray}
\Psi_{e_{+}}^{i,j}&=\frac{1}{\sqrt{2\pi \hbar v_{e_{+}}^{j}}}(j\uu_{-}e^{i \frac{\phi_i}{2}},0,0,\vv_{-}e^{-i \frac{\phi_i}{2}})^{T}e^{i k_{e_{+}}^{j}x}\nonumber\\
\Psi_{e_{-}}^{i,j}&=\frac{1}{\sqrt{2\pi \hbar v_{e_{-}}^{j}}}(0,-j\uu_{+}e^{i \frac{\phi_i}{2}},\vv_{+}e^{-i \frac{\phi_i}{2}},0)^{T}e^{i k_{e_{-}}^{j}x} ,
\label{Eigenfunctions}
\end{eqnarray}
where 
\begin{alignat}{2}
  \uu_\pm  =\sqrt{\frac{\Delta}{2 \epsilon_{\pm}}} e^{\frac{1}{2}h(\epsilon_{\pm})};\quad&\vv_{{\pm}}  = \sqrt{\frac{\Delta}{2 \epsilon_{\pm}}} e^{-\frac{1}{2} h(\epsilon_{\pm})}\nonumber
\end{alignat}
with $\epsilon_{\pm}=\epsilon\pm \epsilon_{DS}(\Phi)$ and  $h(\epsilon_{\pm})= \arccosh{(\epsilon_{\pm}/\Delta)}$ for $\epsilon_{\pm}>\Delta$ and $h(\epsilon_{\pm})=i \arccos{(\epsilon_{\pm}/\Delta)}$ for $\epsilon_{\pm}<\Delta$.
Here, the quasiparticle momentum is $k_{e_\pm}^{j}=\pm k_{F}(j \sqrt{(\epsilon_{\mp}^2-\Delta^2)/\mu^2}+ 1)$ and $v_{e_\pm}^{j}=\hbar^{-1}|\partial_k E_{\pm}^j|=v_F(\uu_{\mp}^2-\vv_{\mp}^2)$ is the associated group velocity. 
The quasihole (QH) eigenfunctions $\Psi_{h_{\pm}}^{i,j}$ can be obtained by replacing $(\uu_\pm,\vv_\pm)\rightarrow (\vv_\pm,\uu_\pm)$, $k_{e_\pm}^{j}\rightarrow k_{h_\mp}^{j}= k_{e_\pm}^{-j}$ and $v_{e_\pm}^{j}\rightarrow v_{h_\mp}^{j}=v_{e_\pm}^{j}$ in the expressions $\Psi_{e_{\mp}}^{i,j}$ of Eq.~(\ref{Eigenfunctions}). Finally, the energy independent tunnel coupling between the N probe and the edge states is described with a symmetric beamsplitter in terms of a spin-independent transmission amplitude $t$. Due to the helical nature of the TI, electrons injected through the probe with spin component collinear with the natural spin quantization axis of the TI edge propagate in one direction, while electrons with opposite spin component propagate in the other one. 

Transport properties of this multiterminal system are determined using the scattering matrix formalism~\cite{sm,datta95,lambert1998}. Our main focus here is the charge current flowing in the probe $J_N^0$ and the heat current $J_{S_L}^1$ flowing in the left superconducting lead $\rm{S_L}$, in response to a small temperature gradient $\delta T$ between the two superconductors $T_{\rm{S_L/S_R}}=T \pm \delta T/2$ and a voltage bias $V_N$ applied to the N probe at temperature $T_{\rm{N}}=T$.
These currents can be written as~\cite{lambert1998}
\begin{eqnarray}
\label{eq:current}
J_i^{k}=\frac{2}{h}\sum_j\sum_{\alpha,\beta}(\alpha e)^{1-k}\int_{0}^{\infty}d\epsilon~(\epsilon-\mu_i)^k\nonumber\\
\times \left(f_i^{\alpha}(\epsilon)-f_j^{\beta}(\epsilon)\right)P_{i,j}^{\alpha,\beta}(\epsilon) ,
\end{eqnarray}
where $k=0$ (only if $i=N$) stands for charge and $k=1$ for heat component and $\alpha, \beta = \pm $ for  QPs and QHs, respectively.
The Fermi functions of the leads $j=\rm{S_L,S_R,N}$ are $f_j^{\alpha}(\epsilon)=\{\exp[(\epsilon-\alpha \mu_j)/k_B T_j)]+1\}^{-1}$, where $\mu_N=eV_N$, $\mu_{\rm{S_L}}=\mu_{\rm{S_R}}=0$, i.e. superconductors are grounded. 
The scattering coefficients $P_{i,j}^{\alpha,\beta}$ represent the reflection ($i=j$) or transmission ($i\neq j$) probabilities of
a quasi-particle of type $\beta$ in lead $j$ to a quasi-particle of type $\alpha$ in lead $i$~\cite{sm,lambert1998}.
It is worth to notice that from their explicit expressions there is no dependence of the scattering coefficients on the contact potential parameter $\Lambda$. This is a direct consequence of the helicity of the edge channels which do not admit ordinary reflections at the interfaces (i.~e.~Klein tunnelling~\cite{lee2019}).
As a further remark, we point out that the scattering approach we are using automatically includes the effects of the Andreev bound states (ABSs) which are sub-gap particle-hole resonant states that localize in the junction.

{\em Results.}--- We now demonstrate and quantify the appearance of a nonlocal thermoelectric response due to the presence of a DS and the helical nature of the topological Josephson junction.
\begin{figure}
	\centering 
	\includegraphics[width=1\columnwidth]{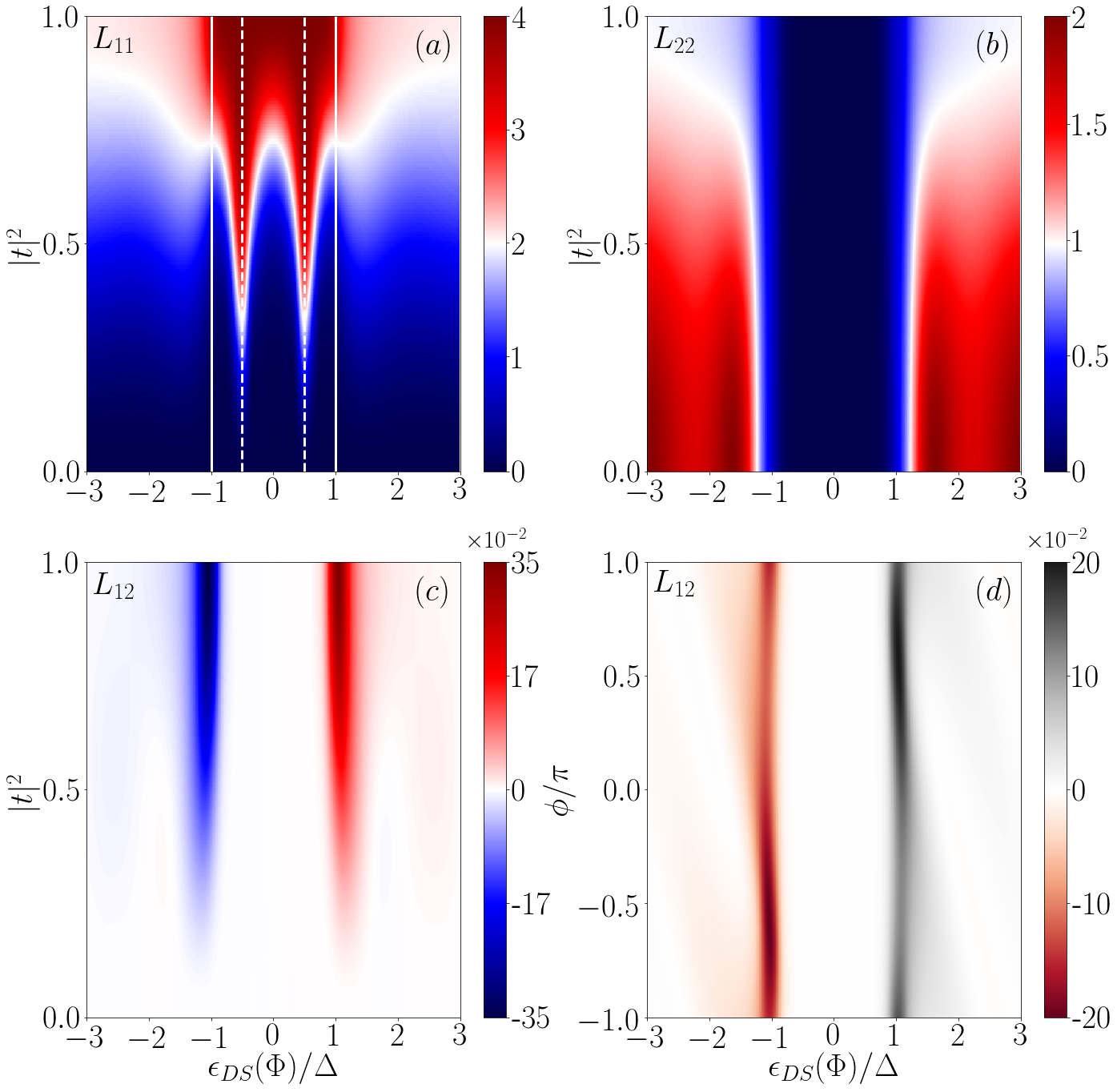}
	\caption{Onsager coefficients $L_{11}$ (a), $L_{22}$ (b) and $L_{12}=-L_{21}$ (c)
	as functions of $\epsilon_{DS}(\Phi)/\Delta$ and the coupling parameter $|t|^2$ for phase bias $\phi=\phi_{\rm{S_L}}-\phi_{\rm{S_R}}=0$, $T/T_C=0.1$ and $L/\xi =1$. (d) $L_{12}$ as a function of  $\epsilon_{DS}(\Phi)/\Delta$ and the phase difference $\phi$ for $\abs{t}^2=0.5$. Such quantities are normalized as follows: $L_{11}/(G_0 T)$, $L_{22}/(G_T T^2)$ and $L_{12}/(\sqrt{G_0 G_T T^3})$.}
	\label{fig:fig2}
\end{figure}
We focus on linear response with $V_N,\delta T\to0$.
A simple physical picture of the thermoelectric mechanism can be grasped analyzing the dispersion-curves 
in Fig.~\ref{fig:fig1}$(b)$. When $\epsilon_{DS}>0$, left-moving (right-moving) QPs $e_{-}$ (QHs $h_{+}$) shift down with respect to the right-moving (left-moving) QPs $e_{+}$ (QHs $h_{-}$). For simplicity, we assume $0<(\Delta-\epsilon_{DS}(\Phi))\sim k_B T\ll (\Delta+\epsilon_{DS}(\Phi))$ so that only left-moving QPs (right-moving QHs), thermalizing with the $T_{\rm{S_R}}$ ($T_{\rm{S_L}}$), contribute to the current. 
This unbalance between the fluxes of cold QPs and hot QHs [see Fig.~\ref{fig:fig1}$(c)$] leads to a thermoelectric current flowing in the N probe.
Moreover, it is worth to notice that, also in the non-linear regime, the thermoelectric current does not depend explicitly on the probe's temperature as long as $T_N\lesssim T$ in order to preserve the physical conditions of the device~\cite{sm}.
In this respect, the fact that $T_N$ is assumed to be exactly in between the temperatures of the two superconductors (see Fig.~\ref{fig:fig1}$(a)$) it is not a necessary requirement.

In addition to the thermoelectric current, a $\Phi$-controlled nonlocal Peltier cooling may be also induced due to the application of a voltage $V_{\rm{N}}$. In this case, as shown in Fig.~\ref{fig:fig1}$(d)$, a charge current from the probe induces mainly left-moving QPs $e_{-}$ and right-moving QHs $h_{+}$ which determine a net energy transport from right to left between the two superconductors even if they are kept at the same temperature. 
Notably, both the sign of the net thermoelectric current and the direction of the cooling can be varied by changing  $\Phi\rightarrow-\Phi$.  These conclusions are not affected by the ABS, since they  do not contribute neither to the thermal nor to the thermoelectrical transport processes.

Quantitatively, the linear response regime is characterized by the following relations~\cite{benenti,taddeibenenti,roura,Hussein,Rafa,Lesovik}
\begin{align}
\label{Osanger}
J_N^0&=L_{11} (V_N/T) + L_{12}(\delta T/T^2)\nonumber\\
J_{S_L}^1&=L_{21} (V_N/T) + L_{22}(\delta T/T^2),
\end{align}
where $V_N/T$ and $\delta T/T^2$ are the two relevant thermodynamic forces (affinities) for the configuration of interest.
Notice that, although the configuration contains three terminals, the driving affinities are two. Hence, the Onsager matrix is effectively $2\times2$~\cite{taddeibenenti,benenti,chiral,roura,mani2018}. Remarkably, in the present setup, the off-diagonal coefficients are nonlocal and satisfy the relation $L_{12}=-L_{21}$.
The behavior of the Onsager coefficients $L_{ij}$ ($i,j=1,2$) are shown in Fig.~\ref{fig:fig2} as functions of 
$\epsilon_{DS}(\Phi)/\Delta$.
The diagonal and local coefficients $L_{11}$ and $L_{22}$ are plotted in units of $G_0T$ and $G_TT^2$, while the nonlocal thermoelectrical coefficient $L_{12}$ is plotted in units of $\sqrt{G_0 G_T T^3}$; with $G_0=2e^2/h$ and $G_T=(\pi^2/3h)k_B^2 T$ being respectively the electrical conductance quantum and the thermal conductance quantum.
In these plots, the length of the junction $L$ is set equal to the superconducting coherence length $\xi=\hbar v_F / \Delta$. Similar results can be obtained in the case of short $L\ll\xi$ or long $L\gg\xi$ junctions~\cite{sm}. 
In Figs.~\ref{fig:fig2}$(a)$ and $(b)$ we plot $L_{11}$ and $L_{22}$, respectively, as functions of $\epsilon_{DS}(\Phi)/\Delta$ and $|t|^2$, setting $\phi=\phi_{\rm{S_L}}-\phi_{\rm{S_R}}=0$.
When the gap is open ($|\epsilon_{DS}(\Phi)|/\Delta<1$), and for low coupling $|t|^2\ll1$, the electrical conductance $L_{11}/(G_0T)$ is almost zero apart from two sharp resonances located at $\epsilon_{DS}(\Phi)/\Delta=\pm 1/2$, where the ABSs cross zero-energy (indicated by white dashed lines in Figs.~\ref{fig:fig2}$(a)$) as expected in the tunneling limit~\cite{ther3}.
By increasing the coupling $|t|^2$ the resonances are broadened as a consequence of the enhancement of the effective linewidth of the ABSs.
When $|t|^2$ increases towards unity, the ABSs are spread and give rise to a finite electrical conductance in the whole range of values of $\epsilon_{DS}(\Phi)$, something that cannot be caught with a tunneling approach.
For all values of $\abs{t}^2$ the thermal conductance $L_{22}$ takes the largest values when the gap is closed $|\epsilon_{DS}(\Phi)|/\Delta>1$, as one can see in Fig.~\ref{fig:fig2}$(b)$.
This is consistent with the fact that in the linear response regime the heat transport in a superconductor is mediated by quasiparticles~\cite{ther1,ther2}.
On the other hand, $L_{22}$ vanishes within the gap when $|\epsilon_{DS}(\Phi)|/\Delta<1$.
This is due to the fact that ABSs cannot allow any thermal transport, while mediating the transport of charge through the Andreev reflection mechanism.
When the gap is closed, the thermal conductance $L_{22}$ presents small fluctuations as a consequence of interference effects and decreases at increasing coupling strength with the probe.
In Fig.~\ref{fig:fig2}$(c)$ we plot $L_{12}$ as a function of $\epsilon_{DS}(\Phi)/\Delta$ and $|t|^2$, with $\phi=0$. We distinguish two peaks at $|\epsilon_{DS}(\Phi)| \sim \Delta $.
This is because in this condition the top left band (for $\epsilon_{DS}(\Phi)\sim\Delta$) and the top right band (for $\epsilon_{DS}(\Phi)\sim-\Delta$) shown in Fig.~\ref{fig:fig1}$(b)$ nearly touch zero energy, thus allowing a small temperature bias to drive a charge current even for a temperature $k_B T\ll \Delta$.
The absolute value of $L_{12}$ increases as a function of $|t|^2$ and its sign changes when changing the sign of DS (or $\Phi$). 
Fig.~\ref{fig:fig2}$(d)$ visualizes the impact of the Josephson phase $\phi$ (vertical axes) in the behavior of the nonlocal thermoelectric coefficient $L_{12}$ for $\abs{t}^2=0.5$.
Here, we can notice that due to symmetry
reasons $L_{12}(\Phi,\phi) \rightarrow -L_{12}(-\Phi,-\phi)$. As a final remark, when $\abs{t}^2\approx1$ (i.~e.~perfect coupling with the probe) $L_{12}$ does not depend neither on the phase bias $\phi$ nor on the junction length $L$.

\begin{figure}[ht!]
	\centering
	\includegraphics[width=1\columnwidth]{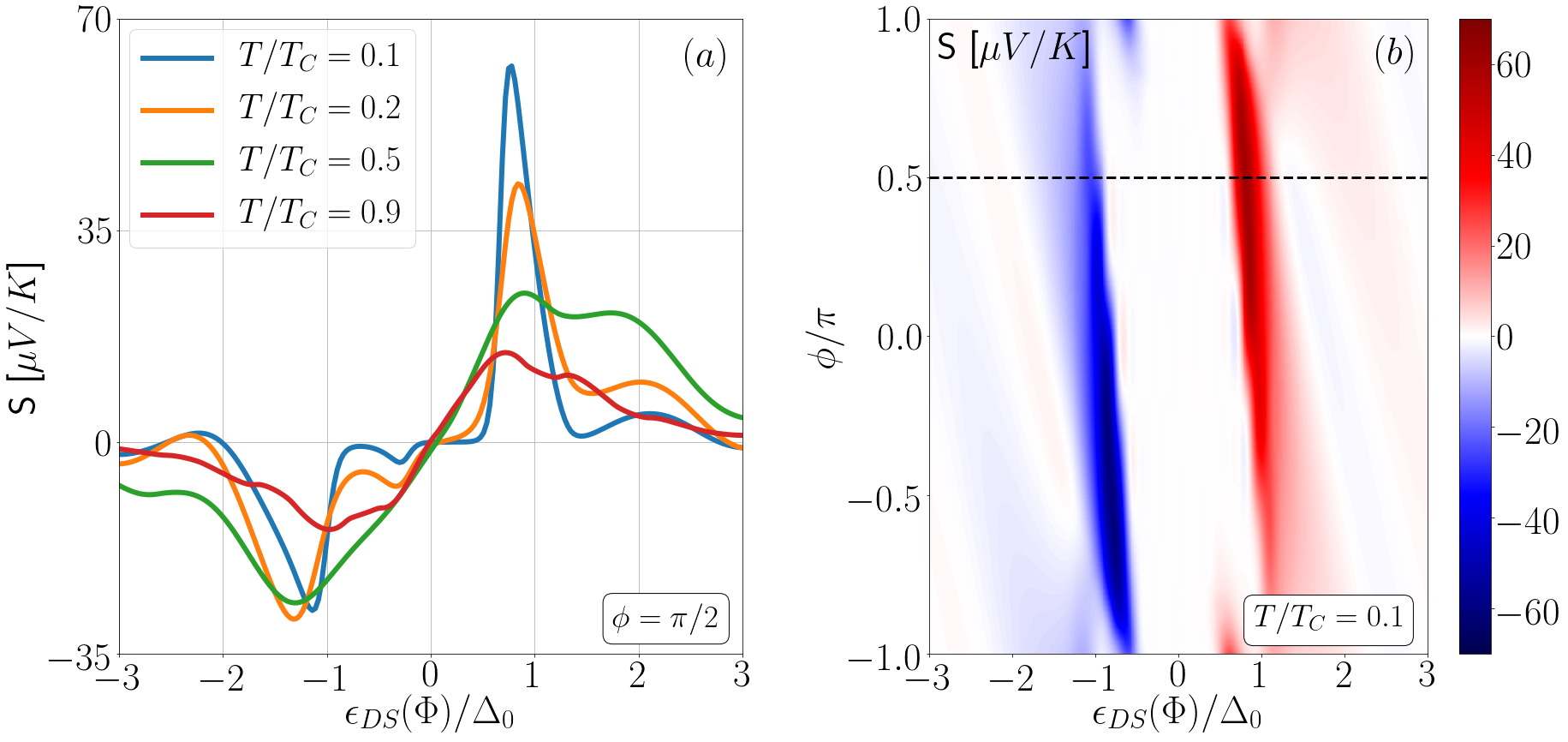}
	\caption{Seebeck coefficient as function of $\epsilon_{DS}(\Phi)/\Delta_0$ for different temperatures at $\phi=\pi/2$~$(a)$ and as function  of $\epsilon_{DS}(\Phi)/\Delta_0$ and $\phi$ for $T/T_C=0.1$~$(b)$. 
	The blue curve in panel $(a)$ correspond to the cut at $\phi=\pi/2$ of the Seebeck coefficient depicted in panel $(b)$ (dashed line) obtained for the same set of parameters: $L/\xi =1$ and $\abs{t}^2=10^{-2}$.}
 \label{fig:fig3}
\end{figure}

To  characterize the nonlocal effect induced by the DS we analyze the nonlocal Seebeck coefficient  $S=(1/T)L_{12}/L_{11}$~\cite{benenti}.
The latter is shown in Fig.~\ref{fig:fig3}, in units of $\mu V/K$, in the case of a weak coupling $|t|^2=10^{-2}$, where the Seebeck coefficient takes the highest values.
In order to make realistic predictions in a wide temperature range, we have also included the self-consistent temperature behavior $\Delta(T)=\Delta_0 \tanh(1.74 \sqrt{T_C/T-1})$, accurate better than $2\%$ with respect to the self-consistent BCS result~\cite{Tinkham,sothmannQD}. 
In Fig.~\ref{fig:fig3}$(a)$ the Seebeck coefficient is reported at $\phi=\pi/2$ for different values of temperatures: its peak value is quite high ($\sim65\mu V/K$), reaching the same order of magnitude of the values predicted for hybrid ferromagnetic-superconducting junctions \cite{seba2,Franz}. 
The maximum value of the Seebeck coefficient decreases by increasing the temperature $T$ and it is reached at $|\epsilon_{DS}(\Phi)|\sim\Delta(T)$. 
The shape of $S$ also depends on  the phase bias $\phi$~(see Fig.~\ref{fig:fig3}$(b)$); namely for $\phi\neq 0$ it is not  antisymmetric $S(\phi,\Phi)\neq -S(\phi,-\Phi)$ with respect to $\Phi$ while it becomes exactly antisymmetric for $\phi=0$.
For completeness, we analyze in the supplemental material~\cite{sm} the figure of merit  $ZT$.
Remarkably, it reaches its maximum value for almost perfect coupling to the probe.

{\em Conclusions.}--- We have discussed a striking consequence of the helical properties of the edge states in a topological Josephson junction in the presence of a normal metal probe coupled to one edge of a quantum spin Hall system. We showed that a thermal gradient between the superconductors in the presence of the Doppler shift generates a nonlocal thermoelectrical transport in the probe even in absence of any spin polarization. By using scattering matrix approach, we have quantitatively evaluated both local and nonlocal Onsager transport coefficients as a function of Doppler shift and phase difference. The nonlocal Seebeck coefficients can achieve high values, comparable with the best hybrid devices based on ferromagnetic elements, in the weak coupling limit (tunneling regime). These nonlocal features are a consequence of the spin-momentum locking of helical states and the induced Doppler shift which can be tuned by means of small external magnetic fields.  
This additional knob can be used to tune the sign of the off-diagonal Onsager coefficient, and therefore to control heat and thermoelectric response in a topological Josephson junction based device.
Such effects are not limited to the tunneling regime, but occur also for an ohmic contact with the probe, provided that the Josephson coupling is not spoiled.
The present device is a very promising tool for probing the helical nature of the edge states in systems where the Hall bar configuration of edge states is difficult to realize.

As a final remark, we notice that our analysis is not limited to the case of full proximization due to a perfect contact, which can be  realized with superconducting pads laying on the TI edges over a few $\mu$m (being much bigger than the superconducting coherence length in the proximized TI $\xi\approx600$ nm)~\cite{expj0,expj1,expj2}. 
A bad contact, though, can be taken into account in our calculations by considering a reduced gap, without changing our results~\cite{sm,Bocquillon_2018_gap}.
Furthermore, a length of the junction $L \sim \xi$ is sufficient to host the contact with a metallic probe (such as an STM tip with a width of 100 nm),  and preserves, at the same time, the ballistic nature of the transport along the edges~\cite{Lunczer_2019_ballistic}. Moreover, studies on the impact of electron-phonon interaction~\cite{eph} and spin-phonon interaction~\cite{sph} in helical edge states support the idea that the transport is ballistic at the operating temperatures for our the setup, typically of a few K.


\begin{acknowledgments}
 We acknowledge support from CNR-CONICET cooperation program “Energy conversion in quantum nanoscale hybrid devices”.  We are sponsored by PIP-RD 20141216-4905 
of CONICET,  PICT-2014-2049 and PICT-2017-2726 from Argentina, as
well as the Alexander von Humboldt
Foundation, Germany and ICTP-Trieste and Simons Foundation (L. A.). A.B. and F.T acknowledge SNS-WIS joint lab QUANTRA. M. C. is supported by the Quant-Era project ``Supertop''. A. B. acknowledge the Royal Society through the International Exchanges between the UK and Italy (Grants No. IES R3 170054 and No. IEC R2 192166).
\end{acknowledgments}



\begin{thebibliography}{9}

\bibitem{moore2009}J. Moore, Nature Phys {\bf 5}, 378–380 (2009)

\bibitem{hasan2010} M. Z. Hasan and C. L. Kane
Rev. Mod. Phys. 82, 3045 (2010)

\bibitem{Qi2010}X-L. Qi, and S-C. Zhang, Rev. Mod. Phys. {\bf 83}, 1057 (2011).

\bibitem{ando2013}Y. Ando, Journal of the Physical Society of Japan {\bf 82} 102001 (2013)

\bibitem{Tkachov_book} G. Tkachov, Topological Insulators: The Physics of Spin Helicity in Quantum Transport (Pan Stanford, Singapore, 2015).

\bibitem{ti1}C. L. Kane and E. J. Mele, Phys. Rev. Lett. {\bf 95}, 146802 (2005).

\bibitem{ti2}C. L. Kane and E. J. Mele, Phys. Rev. Lett. {\bf 95}, 226801 (2005).

\bibitem{ti3}B. A. Bernevig, T. L. Hughes, and S.-C. Zhang, Science {\bf 314}, 1757 (2006).

\bibitem{ti4} M. K\"onig, S. Wiedmann, C. Br\"une, A. Roth, H. Buhmann, L. W. Molenkamp, X. L. Qi, and S-C. Zhang, Science {\bf 318}, 766 (2007).

\bibitem{ti5}
A. Roth, C. Br\"une, H. Buhmann, L. W. Molenkamp, J. Maciejko, X-L. Q, S-C Zhang, Science  {\bf 325}, 294 (2009).

\bibitem{ti6} 
C. Br\"une, A. Roth, H. Buhmann, E. M. Hankiewicz, L. W. Molenkamp, J. Maciejko, X-L. Qi, and S-C. Zhang, Nature Phys. {\bf 8}, 485 (2012).

\bibitem{cob} 
Y. Shi, J. Kahn, Ben Niu, Zaiyao Fei, Bosong Sun, Xinghan Cai, Brian A. Francisco, Di Wu, Zhi-Xun Shen, Xiaodong Xu, David H. Cobden, and Yong-Tao Cui, Science Adv. {\bf 5} (2019)

\bibitem{wu}
S. Wu, V. Fatemi,Q. D. Gibson, K. Watanabe, T. Taniguchi, R. J. Cava, P. Jarillo-Herrero, Science {\bf 359}, 76 (2018).

\bibitem{tang}
S. Tang, C. Zhang, D. Wong, Z. Pedramrazi, H-Z Tsai, C. Jia, B. Moritz, M. Claassen, H. Ryu, S. Kahn, J. Jiang, H. Yan, M. Hashimoto, D. Lu, R. G. Moore, C-C. Hwang, C. Hwang, Z. Hussain, Y. Chen, M. M. Ugeda, Z. Liu, X. Xie, T. P. Devereaux, M. F. Crommie, S-K. Mo, amd Z-X. Shen, Nat. Phys. {\bf 13}, 683
(2017).

\bibitem{jia}Z-Y. Jia, Y-H. Song, X-B. Li, K. Ran, P. Lu, H-J. Zheng, X-Y. Zhu, Z-Q. Shi, J. Sun, J. Wen, D. Xing, and S-C. Li, Phys. Rev. B {\bf 96}, 41108 (2017).

\bibitem{claessen1}
F. Reis, G. Li, L. Dudy, M. Bauernfeind, S. Glass, W. Hanke, R. Thomale, J. Sch\"afer, R. Claessen,
Science {\bf 357}, 21 (2017).

\bibitem{claessen2}G. Li, W. Hanke, E. M. Hankiewicz, F. Reis, J. Sch\"afer, R. Claessen, C. Wu, R. Thomale, Phys. Rev. B {\bf 98}, 165146 (2018).

\bibitem{fuhrer} 
C. Liu, D. Culcer, M. T. Edmonds, M. S. Fuhrer, arXiv:1906.01214

\bibitem{citro2011}R. Citro, F. Romeo, and N. Andrei
Phys. Rev. B {\bf 84}, 161301 (2011)


\bibitem{ronetti2016} F. Ronetti, L. Vannucci, G. Dolcetto,  M. Carrega, M. Sassetti, Phys. Rev. B {\bf 93}(16), 165414 (2016).

\bibitem{ronetti2017} F. Ronetti ,  M. Carrega, D. Ferraro, J. Rech,  T. Jonckheere, T. Martin,  M. Sassetti, Phys. Rev. B, {\bf 95}(11), 115412 (2017).




\bibitem{barbarino2013} S. Barbarino , R. Fazio, M. Sassetti,  F. Taddei, New Journal of Physics, {\bf 15}, 085025 (2013)

\bibitem{teoj1}L. Fu and C. L. Kane, Phys. Rev. B {\bf 79}, 161408(R) (2009).


\bibitem{teoj2}F. Dolcini, M. Houzet, and J. S. Meyer, Phys. Rev. B {\bf 92}, 035428 (2015)

\bibitem{teoj3}S-P. Lee, K. Michaeli, J. Alicea, and A. Yacoby, Phys. Rev. Lett. {\bf 113}, 197001 (2014)

\bibitem{teoj4}G. Tkachov, Phys. Rev. B {\bf 95}, 245407  (2017)

\bibitem{teoj5} G. Blasi, F. Taddei, V. Giovannetti, and A. Braggio, Phys. Rev. B {\bf 99}, 064514 (2019)

\bibitem{teoj6} Tkachov, G., Burset, P., Trauzettel, B., Hankiewicz, E. M., Phys. Rev. B {\bf 92}, 045408  (2015)

\bibitem{kopnin}N. B. Kopnin and A. S. Melnikov
Phys. Rev. B {\bf 84}, 064524 (2011)

\bibitem{tkachovhan}G. Tkachov and E. M. Hankiewicz,
Phys. Rev. B {\bf 88}, 075401 (2013)


\bibitem{ther1}B. Sothmann and E. M. Hankiewicz,
Phys. Rev. B {\bf 94}, 081407 (R), (2016).


\bibitem{ther3}L. Bours, B. Sothmann, M. Carrega, E. Strambini, E. M. Hankiewicz, L. W. Molenkamp, and F. Giazotto, Phys. Rev. Applied {\bf 10}, 014027  (2018).

\bibitem{ther2}
B. Sothmann, F. Giazotto and E. M. Hankiewicz,
New J. Phys. {\bf 19} 023056 (2017).

\bibitem{ther4}D. S. Shapiro, D. E. Feldman, A. D. Mirlin, and A. Shnirman, Phys. Rev. B {\bf 95}, 195425 (2017).

\bibitem{ther5} L. Bours, B. Sothmann, M. Carrega, E. Strambini, A. Braggio, E. M. Hankiewicz, L. W. Molenkamp, and F. Giazotto, Phys. Rev. Applied {\bf 11} 044073 (2019).

\bibitem{expj0} S. Hart, H. Ren, T. Wagner, P. Leubner, M. M\"uhlbauer, C. Br\"une, H. Buhmann, L. W. Molenkamp and A. Yacoby Nature Phys. \textbf{10}, 638 (2014).

\bibitem{expj01} V. S. Pribiag, A. J. A. Beukman, F. Qu, M. C. Cassidy, C. Charpentier, W. Wegscheider and L. P. Kouwenhoven Nature Nanotech. \textbf{10}, 593 (2015).

\bibitem{expj1} E. Bocquillon, R. S. Deacon, J. Wiedenmann, P. Leubner, T. M. Klapwijk, C. Br\"une, K. Ishibashi, H. Buhmann, and L. W. Molenkamp, Nature Nanotech. {\bf 12}, 137 (2017).

\bibitem{expj2} J. Wiedenmann, E. Bocquillon, R. S. Deacon, S. Hartinger, O. Herrmann, T. M. Klapwijk, L. Maier, C. Ames, C. Br\"une, C. Gould, A. Oiwa, K. Ishibashi, S. Tarucha, 
H. Buhmann, and L. W. Molenkamp, Nature Comm. {\bf 7}, 10303 (2016).

\bibitem{das2011} S. Das, S. Rao, Phys. Rev. Lett. {\bf 106}, 236403  (2011).

\bibitem{liu2015} L. Liu, A. Richardella, I. Garate, Y. Zhu, N. Samarth, and C. T. Chen, Phys. Rev. B, {\bf 91}, 235437 (2015)

\bibitem{hus2017} S. M. Hus, X. G. Zhang, G. D. Nguyen, W. Ko,  A. P. Baddor, Y. P. Chen, and A. P. Li,  Phys. Rev. Lett. {\bf 119}, 137202 (2017)

\bibitem{voigtlander2018} B. Voigtl\"ander, V. Cherepanov, S. Korte,  A. Leis, D. Cuma,  S. Just, and  F. L\"upke, Review of Scientific Instruments, {\bf 89}, 101101 (2018).

\bibitem{seba1} 
A. Ozaeta, P. Virtanen,  F. S. Bergeret,  and T. T. Heikkil\"a, Phys. Rev. Lett. {\bf 112}, 057001 (2014).

\bibitem{seba2}F. Sebastian Bergeret, M. Silaev, P. Virtanen, and T. T. Heikkil\"a, Rev. Mod. Phys. {\bf 90}, 041001 (2018).

\bibitem{keidel}
F. Keidel, S-Y. Hwang, B. Trauzettel, B. Sothmann, P. Burset, arxiv:1907.00965

\bibitem{marchegiani} G. Marchegiani, A. Braggio, and F. Giazotto, Phys. Rev. Lett. \textbf{124}, 106801 (2020).

\bibitem{sm} See Supplemental Material at \href{http://link.aps.org/
supplemental/10.1103/PhysRevLett.124.227701}{http://link.aps.org/
supplemental/10.1103/PhysRevLett.124.227701} for details of the calculations and complementary results, which includes Refs. \cite{Ref_sup_6_Giazotto_Pekola_2009,Ref_sup_12_Bernevig_book_2013,Ref_sup_13_Molenkamp_book_2013}.

\bibitem{Ref_sup_6_Giazotto_Pekola_2009} F. Giazotto, T. T. Heikkil\"a, A. Luukanen, A. M. Savin, and J. P. Pekola
Rev. Mod. Phys. \textbf{78}, 217 (2009)

\bibitem{Ref_sup_12_Bernevig_book_2013} B. A. Bernevig, T. L. Hughes. \textit{Topological insulators and topological superconductors}. Princeton university press, 2013.

\bibitem{Ref_sup_13_Molenkamp_book_2013} M. Franz, L. Molenkamp. \textit{Topological Insulators}. Elsevier, 2013.

\bibitem{datta95}S. Datta, \textit{Electronic Transport in Mesoscopic Systems} (Cambridge University Press, Cambridge, England, 1995).

\bibitem{lambert1998}C. J. Lambert and R. Raimondi, J. Phys. Condens. Matter {\bf10}, 901 (1998)



\bibitem{lee2019}S. Lee, V. Stanev, X. Zhang et al., Nature {\bf 570}, 344 (2019) 







\bibitem{taddeibenenti}F. Mazza, R. Bosisio, G. Benenti, V. Giovannetti, R. Fazio and F. Taddei, New J. Phys. {\bf 16}, 085001 (2014).

\bibitem{Rafa}  R. S\'anchez, P. Burset, and A. Levy Yeyati, Phys. Rev. B {\bf 98}, 241414(R) (2018)

\bibitem{Hussein} R. Hussein, M. Governale, S. Kohler, W. Belzig, F. Giazotto, and A. Braggio, Phys. Rev. B {\bf 99}, 075429 (2019)


\bibitem{Lesovik} N. S. Kirsanov, Z. B. Tan, D. S. Golubev, P. J. Hakonen, and G. B. Lesovik, Phys. Rev. B {\bf 99}, 115127 (2019) 

\bibitem{roura}P. Roura-Bas, L. Arrachea, and E. Fradkin, Phys. Rev. B {\bf 98}, 195429 (2018).



\bibitem{benenti}G. Benenti, G. Casati, K. Saito, R. S. Whitney, Phys. Rep. {\bf 694}, 1 (2017).

\bibitem{chiral} R. S\'anchez, B. Sothmann, A. N. Jordan, Phys. Rev. Lett. {\bf 114}, 146801 (2015).

\bibitem{mani2018}A. Mani, C. Benjamin, Phys. Rev. E {\bf 97}(2), 022114 (2018).

\bibitem{sothmannQD}  M. Kamp, B. Sothmann, Phys. Rev. B {\bf 99}, 045428 (2019) 

 \bibitem{Tinkham} Tinkham, M., \emph{Introduction to superconductivity} (McGraw-Hill, New York, 1996)
 







\bibitem{Franz} F. Giazotto, P. Solinas, A. Braggio, and F.S. Bergeret, Phys. Rev. Applied 4, 044016 (2015).

\bibitem{Bocquillon_2018_gap}  E. Bocquillon, J. Wiedenmann, R. S. Deacon, T. M. Klapwijk, H. Buhmann, L. W. Molenkamp (2018) \textit{Topological Matter}. Springer, Cham.

\bibitem{Lunczer_2019_ballistic} L. Lunczer, P. Leubner, M. Endres, V. L. Müller, C. Br\"une, H. Buhmann, and L. W. Molenkamp, Phys. Rev. Lett. {\bf 123}, 047701 (2019)


\bibitem{eph}J. C. Budich, F. Dolcini, P. Recher, and B. Trauzettel, Phys. Rev. Lett. {\bf 108}, 086602 (2012)

\bibitem{sph}S. Groenendijk, G. Dolcetto, and T. L. Schmidt, Phys. Rev. B {\bf 97}, 241406 (2018)


%
%
%
%
%
%
%
%
  

 

\end{thebibliography}
\end{document}